\documentclass[prl,superscriptaddress,aps,twocolumn,amsmath,nofootinbib]{revtex4}
\usepackage{epsfig}
\usepackage{graphics}
\usepackage{color}

\def\prl{{ Phys. Rev. Lett.}~}

\def\rmp{{ Rev. Mod. Phys.}~}

\def\etal{{ et al.}}

\newcommand{\vecr}{\mathbf{r}}

\newcommand{\vecn}{\mathbf{n}}

\newcommand{\vecE}{\mathbf{E}}

\newcommand{\beq}{\begin{equation}}
\newcommand{\eeq}{\end{equation}}

\begin{document}

\title{Can extreme electromagnetic fields accelerate the $\alpha$ decay of nuclei?}
\author{Adriana P\'alffy}
\affiliation{Max-Planck-Institut f\"ur Kernphysik, Saupfercheckweg 1, 69117, Heidelberg, Germany}
\author{Sergey V. Popruzhenko}
\affiliation{Prokhorov  General Physics Institute of the Russian Academy of Sciences, Vavilov Str. 38, 119991, Moscow, Russia}
\affiliation{Department of Physics, Voronezh State University, Universitetskaya pl.1, 394018, Voronezh, Russia}
\affiliation{Max-Planck-Institut f\"ur Physik komplexer Systeme, N\"othnitzer Strasse 38, 01187, Dresden, Germany}


\begin{abstract}
The possibility to control the $\alpha$ decay channel of atomic nuclei with electromagnetic fields of extreme intensities envisaged for the near future at multi-petawatt and exawatt laser facilities is investigated theoretically.
Using both analytic arguments based on the Wentzel-Kramers-Brillouin  approximation and numerical calculations for the imaginary time method applied in the framework of the $\alpha$ decay precluster model, we show that no experimentally detectable modification of the $\alpha$ decay rate can be observed with super-intense lasers at any so-far available wavelength. Comparing our predictions with those reported in several recent publications, where a considerable or even giant laser-induced enhancement of the decay rate has been claimed, we identify there the misuse of a standard approximation.
\end{abstract}

\maketitle


{\it Introduction.} The upcoming commissioning of new laser sources of few up to  10 petawatt (PW) power  \cite{ELI,J-Karen,SULF,P3-ELI,light-f}, and the plans to build even more powerful sub-exawatt laser systems \cite{light-f,XCELS} have triggered theoretical revisions of a variety of phenomena induced or assisted by electromagnetic fields of extreme intensity. 
This includes in particular the creation of electron-positron pairs by electromagnetic fields from vacuum, laser initiation of quantum electrodynamic cascades of elementary particles and plasma dynamics in the classical or quantum radiation-dominated regime of interaction, see for instance reviews in Refs.~\cite{bulanov-rmp09,dipiazza-rmp12,fedotov-cp15} and references therein.
Within this new field, several theoretical proposals to consider the effect of laser radiation on nuclear processes and in particular on $\alpha$  decay have been recently reported \cite{hector,misicu-jpg13,misicu-degruyter16,delion-prl17,kis-jpg18,bai,qi-prc19}.

The radioactive $\alpha$ and $\beta$ decays of atomic nuclei are most fundamental and common nuclear processes observed in the Universe and in laboratory experiments \cite{BlattWeisskopf,radioactive_decays}.  
The theory of $\alpha$ decay was one of the early successes of quantum mechanics as in 1928 both Gamow \cite{Gamow1928} and independently Condon and Gurney \cite{ConGur1928} used the new concept of tunneling to calculate $\alpha$ decay lifetimes. 
At present, within this tunneling picture  both $\alpha$ and proton radioactivity are related to widths and shifts of quasistationary states in the quantum mechanical two-potential approach \cite{GurKalb1987} accounting for the spectroscopic factor for the proton and the preformation factor of the $\alpha$ particle in the nucleus. 
Intuitively, in order to significantly alter the $\alpha$ decay probability, an external electromagnetic field should be able to considerably change the energy $Q_{\alpha}$ of the escaping $\alpha$ particle on the spatial length determined by the width of the Coulomb barrier that is tunneled, $l_{\alpha}\simeq 10^{-11}$cm, on time scales $\tau_{\alpha}$ similar to the time required for this particle to cross the barrier. 
Considering a generic value of 5 MeV $\alpha$ particle energy,  $\tau_{\alpha}\simeq 10^{-21}$s. 
On this time scale, any  electromagnetic field generated by laboratory sources of intense coherent radiation with photon energies $\hbar\omega$ from 0.1 eV (${\rm CO}_2$ laser) to 10 keV (X-ray Free Electron Lasers) \cite{XFEL} can be considered quasi-static, as its oscillation cycle $T_{\omega}\gg\tau_{\alpha}$.
Then, the energy gained from an external field of strength $E$ is of the order of $\Delta Q_{\alpha}\simeq Z_\alpha e El_{\alpha}$ with $e$ being the elementary charge and $Z_ \alpha=2$ the atomic charge of the $\alpha$ particle, respectively. 
An order of magnitude estimate for the electric field value that would produce a seizable modification of the $\alpha$ decay rate in this picture gives
 $E^*\simeq Q_{\alpha}/(Z_{\alpha} e l_{\alpha})\approx10^{19}$ V/cm.
 This value exceeds the critical field of quantum electrodynamics \cite{sauter,heisenberg,schwinger} $E_{\rm cr}=m_e^2c^3/e\hbar=1.32\cdot 10^{16}\ {\rm V/cm}$ with $m_e$ being the electron mass,  by almost three orders of magnitude and is far beyond present laser capabilities. 
 
Present record intensities of electromagnetic fields created at sub-PW and PW laser facilities  hardly exceed $10^{22}$W/cm$^2$ \cite{yan-oe08,PW-lasers} which corresponds to the electric field strength $E_0\approx 3\cdot 10^{12}$V/cm. 
Such laser fields can considerably modify the spectra of the decay products, since any charged particle promptly appearing in the continuum dressed by electromagnetic waves will be accelerated toward the detector with the momentum gain determined by the instant of release (see e.g. Refs.~\cite{ritus,becker} for $\beta$  and Ref.~\cite{hector} for $\alpha$ decay of nuclei). 
They however have no observable effect on the overall decay probability. 

In a striking contrast to these estimates, several recent theoretical works have predicted a strong effect of laser radiation on $\alpha$ decay even at currently achievable intensities \cite{misicu-jpg13,delion-prl17,kis-jpg18,bai,misicu-degruyter16}. 
These spectacular claims brought super-intense lasers in  discussions about  practical applications for recycling of nuclear waste \cite{rev}. 
This controversy and the appealing applications that are at stake call for a reliable and thorough theoretical investigation of laser-assisted $\alpha$ decay at novel ultra-strong laser facilities. 
In this Letter, we derive a correct value for the external electric field strength which can alter the decay rate using two different approaches, one analytical based on the Wentzel-Kramers-Brillouin (WKB) approximation of quantum mechanics and one numerical employing the imaginary time method (ITM) \cite{popov-usp04,popov-itm} extended for quasi-stationary states of $\alpha$ emitters considered in the precluster model \cite{lat2011}. 
We apply these two approaches to calculate the laser-assisted $\alpha$ decay rates of several  $\alpha$ emitters with lifetimes spanning from $10^{-7}$s to $10^{15}$ years.
The two  methods agree well with each other and demonstrate  that although the value $E^*$ is an overestimate, neither present nor upcoming laser fields can produce any seizable effect on $\alpha$ decay rates. 
We address the controversy with results of Refs.~\cite{delion-prl17,kis-jpg18,bai} and conclude that the gross enhancements in $\alpha$ decay rates predicted there are most likely due to  the misuse of an approximation.


{\it WKB method.} 
In order to analytically quantify the effect of an electromagnetic field on the rate of $\alpha$ decay we employ the Gamow model \cite{Gamow1928}. A preformed $\alpha$ particle tunnels under the barrier generated by the Coulomb force. 
We note that {\bf (i)} the relevant time over which the laser can assist the tunneling process is the time-of-flight $\tau_{\alpha}$ of the $\alpha$ particle under the potential barrier.
This time is very short and does not coincide with the observed $\alpha$-decay lifetimes $t_{1/2}$.
While $\tau_{\alpha}$ always remains shorter than $10^{-20}$s, the lifetimes $t_{1/2}$ span orders of magnitude from ns to the age of the Universe. 
{\bf (ii)} For electromagnetic fields with strength well below $E^*$  the Coulomb forces greatly exceed the Lorentz force acting on the particle in the field of a laser wave. 
Finally, {\bf (iii)} the sub-barrier motion of the $\alpha$ particle remains nonrelativistic even in extremely strong fields, due to the shortness of the escape time $\tau_{\alpha}$ and the large mass of the $\alpha$ particle compared to the one of the electron.  

Thus {\bf (i)} allows us to treat the electric field effect in the quasistatic approximation, while {\bf (ii)} and {\bf (iii)} justify to  calculate laser-induced corrections to the barrier penetrability using a nonrelativistic perturbation theory.
Within the quasistatic approach, time is considered as a parameter, and, when the interaction with the laser field is described in the length gauge and the nuclear distortion of the Coulomb barrier at small distances is discarded, the variables are separable in parabolic coordinates \cite{LandauBook}.
These coordinates should be used to describe correctly the angular dependence of the decay rate \cite{popov-jetpl91}.
However, a reliable estimate can be made within a 1D model assuming the sub-barrier trajectory of the $\alpha$-particle a straight line as in the field-free case. The Gamow decay rate is then (in Gaussian units)
\beq
R\approx \nu_0\exp\left(\mkern-7mu -\frac{2}{\hbar}\int\limits_0^{b} \mkern-9mu \sqrt{2 m_r [V(r)-ez_{\rm eff}\vecE(t)\cdot\vecr-Q_{\alpha}]}dr\right)\, .
\label{P}
\eeq
Here, $\nu_0$ is the frequency of $\alpha$ particle oscillations inside the nucleus \cite{Fermi},  $m_r$ is the reduced mass of the nuclear system composed of $\alpha$ particle and daughter nucleus, and  $V(r)$ is the Coulomb potential the particle tunnels through. Furthermore, 
$\vecE(t)$ is the time-dependent electric field of the laser,
$b=2Ze^2/Q_{\alpha}$ is the barrier exit point, $Z$ and $A$ are the charge and atomic numbers of the daughter nucleus and $z_{\rm eff}=(2A-4Z)/(A+4)$ is the effective charge which accounts for the center of mass motion of the decaying system in the external electromagnetic field.
For electromagnetic fields of amplitude $E_0\ll E^*$ the potential energy in the laser field $\vert ez_{\rm eff}\vecE(t)\cdot\vecr\vert$ is small compared to the absolute value of the kinetic energy $\vert Q_{\alpha}-V(r)\vert$ everywhere except the vicinity of the turning points. 
These however do not make a considerable contribution for the semiclassical integral in Eq.~(\ref{P}) above. 
Using a series expansion of the integrand up to the second order in $E(t)$ and performing the integration we obtain for the laser-induced factor in the rate $R=R_0\cdot R_L$ 
\beq
R_L\approx\exp\left(\frac{2E(t)}{E_{\rm eff}}\vecn\cdot\vecn_0-\frac{35}{9\pi\nu_{\alpha}}\frac{E^2(t)}{E^2_{\rm eff}}(\vecn\cdot\vecn_0)^2\right)
\label{PL}
\eeq
where $R_0=\nu_0\exp(-2\pi\nu_{\alpha})$ is the field-free rate, $\vecn$ and $\vecn_0$ are the unit vectors along the particle emission direction and the laser polarization, respectively, and 
\beq
E_{\rm eff}=\frac{2\hbar \sqrt{2}Q_{\alpha}^{5/2}}{3\pi Z^2z_{\rm eff}e^5\sqrt{m_r}}~,~~~\nu_{\alpha}=\frac{2Ze^2}{\hbar v_{\alpha}}~.
\label{Eeff}
\eeq
Here $\nu_{\alpha}$ is the Sommerfeld parameter for $\alpha$ decay and $v_{\alpha}$ the $\alpha$-particle velocity corresponding to $Q_{\alpha}$.

These simple formulas show that a considerable electromagnetic effect on the rate of  $\alpha$ decay will be achieved already at $E_0\simeq E_{\rm eff}\ll E^*$. 
It is not uncommon for tunneling phenomena that the actual value of the external field which considerably alters the rate appears much smaller than that necessary to strongly affect the value of the particle momentum under the barrier.
As an example, a fast tunneling of atomic electrons in a static external electric field may happen already at field strengths on the level of 5\% of the characteristic atomic field on the respective Bohr orbit \cite{popov-usp04,poprz-pra19}.
In our case, the reason for that  is the numerically large value of the field-free sub-barrier action $2\pi\nu_{\alpha}\gg 1$.
At $E(t)= E_{\rm eff}$ the time-dependent factor $R_L$ enhances the rate by $\approx e^2=7.78$ times.
The characteristic electric field  $E_{\rm eff}$ is a factor $3\pi\nu_{\alpha}/8$ times less than $E^*$, which gives, for $^{232}$Pu with $Z=92$ and $Q_{\alpha}\approx 5$~MeV, $E_{\rm eff}\approx 8\cdot 10^{16}\ {\rm V/cm}\approx 10^{-2}E^*$.
Consequently, expression (\ref{PL}) which was obtained by perturbative expansion is valid for $E\ll E^*$ and therefore applies for electric field strengths $E\simeq E_{\rm eff}$, i.e. when the laser effect on the rate is already numerically large, $R_L\gg 1$. 
At the same time, as long as $E_0\ll E^*$, the second term in the exponent of (\ref{PL}) can be neglected.
We may therefore denote $E_{\rm eff}$ as threshold electric field strength that is sufficient to modify the $\alpha$ decay process, a conclusion which is independent of the laser photon energy in a broad frequency domain $\hbar\omega\ll Q_{\alpha}/\nu_{\alpha}$.
The values of $E_{\rm eff}$ for several $\alpha$-decaying nuclei are shown in Table I; the lowest of them (for $^{144}_{60}\mathrm{Nd}$) corresponds to the intensity $I\approx 5\cdot 10^{29}$W/cm$^2$.
Thus, for practical calculations, the series expansion can be used in Eq.~(\ref{PL}) to find the relative effect of the laser field on the $\alpha$ decay rate, $(R-R_0)/R_0=R_L-1$. Then,  the time-dependent and time- and angle-averaged field-induced factors  in the decay rate, respectively, are given by
\beq
R_L-1\approx\frac{2E(t)}{E_{\rm eff}}\vecn\cdot\vecn_0~,~~~\bar{R}_L-1\approx\frac{E_0^2}{3E_{\rm eff}^2}~.
\label{PL-1}
\eeq
Estimates made along Eqs.~(\ref{PL-1}) are shown in Table \ref{table} for the $\alpha$ emitters
$^{106}_{52}\mathrm{Te}$, $^{144}_{60}\mathrm{Nd}$, $^{162}_{74}\mathrm{W}$, 
$^{212}_{84}\mathrm{Po}$, $^{238}_{94}\mathrm{Pu}$ and $^{238}_{92}\mathrm{U}$ at the laser intensity $I=10^{26}$ W/cm$^2$ which is currently considered as an optimistic upper limit for the experimental achievements expected in a near future \cite{XCELS}.

\begin{table*}[t]
\begin{center}
  \begin{tabular}{ | l | c | c | c | c | c | c | c | c | c | c | c | }
    \hline
   Isotope & $Q_{\alpha}$(MeV) & $c_1$(fm) & $t^{ex}_{1/2}$(s) & $t^{th}_{1/2}$(s) & $E_{\rm eff}$(V/cm) & $R_L-1$ & $\bar{R}_L-1$ & $R'_L-1$ & $N_0$ & $\Delta N$ \\ \hline
$^{106}_{52}\mathrm{Te}$ & 4.325 & 1.486 & $7\cdot 10^{-5}$ & $6.1\cdot 10^{-5}$ & $9.83\cdot 10^{17}$ &  $5.57\cdot 10^{-4}$ & $2.59\cdot 10^{-8}$ & $6.37\cdot 10^{-8}$ & $10^2$ & $3\cdot 10^{-6}$ \\ \hline
$^{144}_{60}\mathrm{Nd}$ & 1.907 & 1.484 &  $7.2\cdot 10^{22}$& $5.6\cdot 10^{22}$ & $2.12\cdot 10^{16}$ & $2.57\cdot 10^{-2}$ & $5.54\cdot 10^{-5}$ & $1.59\cdot 10^{-4}$ & $10^{-25}$ & $5\cdot 10^{-30}$ \\ \hline
$^{162}_{74}\mathrm{W}$ & 5.675 & 1.432   & 1.39  & 2.45 & $4.05\cdot 10^{17}$ & $1.35\cdot 10^{-3}$  &$1.52\cdot 10^{-7}$ & $4.35\cdot 10^{-7}$ & $5\cdot 10^{-3}$ & $8\cdot 10^{-10}$\\ \hline
$^{212}_{84}\mathrm{Po}$ & 8.953 & 1.409   & 2.99$\cdot 10^{-7}$  & 1.6$\cdot 10^{-7}$& $3.88\cdot 10^{17}$ &$1.35\cdot 10^{-3}$  & $1.66\cdot 10^{-7}$&$4.88\cdot 10^{-7}$ &$2\cdot 10^4$& $4\cdot 10^{-3}$\\ \hline
$^{238}_{94}\mathrm{Pu}$ & 5.593 & 1.390   & $2.77\cdot 10^{9}$ & $4.4\cdot 10^{9}$ &  $9.81\cdot 10^{16}$ & $5.58\cdot 10^{-3}$ & $2.60\cdot 10^{-6}$ & $7.69\cdot 10^{-6}$ & $3\cdot 10^{-12}$ & $6\cdot 10^{-18}$ \\ \hline
$^{238}_{92}\mathrm{U}$ & 4.274 & 1.394  & $1.4\cdot 10^{17}$ &  $4.3 \cdot 10^{17}$ & $4.84\cdot 10^{16}$ &$1.13\cdot 10^{-2}$  & $1.06\cdot 10^{-5}$ & $3.16\cdot 10^{-5}$ & $5\cdot 10^{-20}$& $5\cdot 10^{-25}$\\ \hline

\end{tabular}
\end{center}

\caption{Energy of the $\alpha$ particle, parameter $c_1$ of the model potential \cite{Buck},  experimental (\emph{ex}) \cite{ensdf} and theoretical (\emph{th})   half-lives (calculated from the model of \cite{Buck}), and the corresponding field-free decay events $N_0$ in a $\lambda^3$ focal spot for $\lambda=1\ \mu{\rm m}$ during the laser pulse duration $\tau=100$~fs. Results from the WKB model: the characteristic electric field (\ref{Eeff}), the linear ($R_L-1$) and average ($\bar{R}_L-1)$ corrections to the rate (\ref{PL-1}), and the laser-induced change of the number of $\alpha$ decay events $\Delta N=N_0(\bar{R}_L-1)$. Results from ITM: the numerically calculated correction $R'_L-1$ averaged over the laser period. 
The considered laser field strength $E_0=2.74\cdot 10^{14}$V/cm corresponds to $I=10^{26}$W/cm$^2$ and the nuclear potential depth was taken U$_0$=135.6 MeV \cite{Buck}.
\label{table}}
\label{table1}
\end{table*}


{\it ITM in the precluster model.} 
Our second approach  considers the field-assisted tunneling of the preformed $\alpha$ particle through the Coulomb barrier of the nucleus, following  the framework of the phenomenological precluster model \cite{Buck}. 
In this  modification of the Gamow model, deviations of the interaction potential $V(r)$ from the pure Coulomb form at short distances from the nucleus are considered. 
The preformed $\alpha$ cluster is initially confined in a potential well with depth $-\mathrm{U}_0$, which is taken as the mean field nuclear potential that the nucleons of the parent nucleus experience. 
This nuclear potential has a finite (short) length which is given by  x$_0=c_1 A_p^{1/3}$ with $c_1$ a constant that defines the radius of the parent nucleus and $A_p$ the mass number of the latter \cite{Buck,Royer}. 
For distances larger than x$_0$, the  interaction is dictated by the Coulomb force acting between the protons of the daughter nucleus and the $\alpha$ particle.  
Technically this more realistic description for the nuclear potential results in a better estimate for the frequency $\nu_0$ in (\ref{P}) and in the replacement of the lower integration limit there by the potential well radius x$_0\ll b$. 
For field-free decays this model gives a fairly good agreement with experimental data on nuclear halflives, as it is shown in Table \ref{table} where both experimental and theoretical values for  several $\alpha$ emitters are given.

According to the ITM \cite{popov-usp04,popov-itm}, a trajectory x$(t)$ satisfying the Newton equation
\beq
m_r\ddot{\mathrm{x}}=\frac{ZZ_\alpha}{\mathrm{x}^2} + ez_{\rm eff}E(t)\, ,
\eeq
can be found along which the particle starts its motion at the complex time instant $t=t_s$ inside the well, x$(t_s)=$x$_0$, arrives at the exit of the barrier when $t=t_0$ and has at $t\to\infty$ the energy equal to $Q_{\alpha}$. 
The exit point is separated from the well by the classically forbidden region, so that the solution of the Newton equation satisfying the assigned initial conditions only exists in complex time, $t=t_0+i\tau$.  
The tunneling rate is given by \cite{popov-usp04,popov-itm} $R\approx\nu_0\exp(-2{\rm Im}[W])$ where the classical action $W$ is found along the complex trajectories under the barrier, starting from the modified strong-field approximation transition amplitude \cite{lat2011,hector}. 
Also here we use a one-dimensional approach, following  successful models that have proven their predictive power in non-relativistic laser-atom interactions \cite{Eberly}. 
Thus our ITM results correspond to the case of particle emission in the direction of the laser field polarization, $\vecn\cdot\vecn_0=1$.
On the other hand, compared to the WKB model approach presented above, the ITM takes into account numerically the field-induced modifications of the $\alpha$ particle trajectory. 

For an estimate of the laser effect on the $\alpha$ decay rate, we have  considered the idealized case of a monochromatic field with  $E_0=2.74\cdot 10^{14}$~V/cm, corresponding to an intensity $I=10^{26}$~W/cm$^2$, and 800~nm wavelength. 
We average the field-assisted $\alpha$ decay rate over the laser field period which corresponds to the second of Eqs.(\ref{PL-1}) without the factor $1/3$ which comes from the angular integration. 
Comparing the relative rates $R_L-1$, $\bar{R}_L-1$ and $R'_L-1$, shown in Table  \ref{table} we find a very good agreement between the WKB and ITM results. 
Our results are also consistent with previous results in Ref.~\cite{hector} when taking into account that there the charge $Z_\alpha=2$ was used instead  of $z_{\rm eff}$, neglecting the motion of the daughter nucleus in the laser field.

{\it Possible observation.} Eq.~(\ref{Eeff}) and the numerical results in Table  \ref{table} show that the value of the threshold electric field grows with the $\alpha$ particle energy and decreases with the charge $Z$ of the daughter nucleus. 
Therefore, the effect of the laser field on $\alpha$ decay grows exponentially for heavy nuclei with large atomic number $Z_p=Z+2$ and for decays characterized by relatively small values $Q_{\alpha}$.
This behavior is not surprising taking into account that the Coulomb barrier width grows as $b=2Ze^2/Q_{\alpha}$, so that the quasi-static electric field of the laser can make a more significant work during the (longer) tunneling process.
Also the value of $E_{\rm eff}$ appears smaller for heavier nuclei where $z_{\rm eff}$ is larger.
However, these mechanisms of the threshold suppression do not make the effect experimentally detectable, as with decreasing of $Q_{\alpha}$ the field-free penetrability drops down  much faster than $R_L$ grows.  The nucleus becomes then practically stable despite the $\alpha$ decay not being formally forbidden, see for instance the case of $^{144}_{60}\mathrm{Nd}$ in Table \ref{table}  with $Q_{\alpha}=1.9$ MeV, the minimal value $E_{\rm eff}\approx 2.2\cdot 10^{16}$~V/cm, but field-free half-life $7.2\cdot 10^{15}$ years.

Finally, we note that a modification of the instant decay rate by $\simeq 1$\% and that of the averaged one by $\simeq 10^{-5}$ which may happen for some nuclei according to our predictions in Table \ref{table} at intensities $I\simeq 10^{26}$~W/cm$^2$ leaves no chance for experimental detection.
Extreme laser intensities can only be reached under a tight focusing limiting the interaction volume by $\simeq\lambda^3\approx 10^{-12}\ {\rm cm}^3$ for infrared lasers ($\lambda\simeq 1\ \mu{\rm m}$) and by a much smaller volume for X-ray lasers.
Assuming a solid state density target with a generic density of $10^{23}$ cm$^{-3}$, the number of atoms in the interaction volume is $N\simeq 10^{11}$. 
The number $N_0$ of field-free decays per laser shot and its change $\Delta N$ due to the laser effect are shown in the last two columns of Table \ref{table} considering a pulse duration $\tau=100$~fs. 
These numbers show clearly that the laser-induced corrections are for all practical purposes negligible. 
For $^{144}_{60}\mathrm{Nd}$ where the correction (\ref{PL-1}) achieves its maximum, the field-free decay rate is so small that no events will ever practically happen in the laser focus. 
The case of x-ray lasers is even less realistic due to a much smaller volume over which one would need to focus to achieve extreme intensity values. 

{\it Contradicting theoretical predictions.}
We turn now to recent publications where a gross effect of laser fields on $\alpha$ decay has been predicted.
In a Letter \cite{delion-prl17}, Delion and  Ghinescu claimed based on analytical considerations that  $\alpha$-decay can be significantly accelerated in a strong laser field with intensity $I\sim 10^{20}\div 10^{22}$W/cm$^2$. 
For a quantitative analysis, Ref.~\cite{delion-prl17} introduces the dimensionless parameter $D$ which is proportional to the ratio of the quiver amplitude the $\alpha$-particle has in an external electromagnetic field to the nuclear radius $R_0$, $D\propto\sqrt{I}/(\omega^2 R_0)$. 
Ref.~\cite{delion-prl17} claims that for laser fields with $D>1$, a considerable modification of the  $\alpha$-decay rate can be achieved, and 
in particular for $D=3$, this would lead to a six orders of magnitude enhancement for the case of $^{232}$Pu.  
But for a Ti-sapphire laser with $\hbar\omega\approx 1.5$ eV,  $D$ reaches unity already at modest intensities of $I\approx 5\cdot 10^{12}$ W/cm$^2$.
Were the theoretical predictions of Ref.~\cite{delion-prl17} correct, this enhancement by many orders of magnitude in the $\alpha$-decay rate could have been easily achieved with lasers routinely used in labs for the last several decades, or even using a sufficiently powerful microwave oven \cite{misha}.

Upon close inspection, it appears that these surprising predictions  stem from the misuse of an approximation. 
In order to approach the problem of $\alpha$-decay in the presence of an external time-dependent field analytically, Ref.~\cite{delion-prl17} employs the Kramers-Henneberger transformation (KHT) \cite{henneberger} and  proceeds with an essential approximation by replacing the time-dependent Coulomb potential in Eq.~(6) by its static (i.e., averaged over the laser period) component. 
This ansatz is applicable only provided that the  characteristic time of the processes under consideration remains greater (ideally, much greater) than the laser period. 
As this is indeed the case for a broad variety of atomic phenomena in laser fields, KHT is widely used in atomic physics (see e.g. the book \cite{fedorov} and references therein).
However, for the considered problem exactly the opposite condition $T_{\omega}\gg\tau_{\alpha}$ is satisfied, as we have pointed out above.
This  invalidates all results of Ref.~\cite{delion-prl17}.
A similar error appears also in Refs.~\cite{kis-jpg18,bai}.
The predictions on Refs.~\cite{misicu-jpg13,misicu-degruyter16} based on the KHT and a numerical solution of the Schr\"odinger equation are even more counter-intuitive than the ones of Ref.~\cite{delion-prl17,kis-jpg18,bai} and are likely to stem from numerical inaccuracies. 

Finally, calculations by Qi et al. in \cite{qi-prc19} are close to those we have presented here, and the numerical values for the linear correction agree quantitatively with our estimates.
In contrast to our conclusions, the authors expect the effect to be experimentally detectable and suggest to use elliptically polarized radiation to avoid the cancellation of the linear term in (\ref{PL-1}) after the time averaging.
However, the angular averaging of the term $\vecn\cdot\vecn_0$ will lead to the cancelation of the linear contribution irrespectively of the laser polarization state.  Furthermore, the authors of Ref.~\cite{qi-prc19} do not consider the realistic number of nuclei decaying within the small laser focus. 


{\it Conclusions.} By employing two complementary approaches, one analytical based on the WKB approximation and one numerical using the ITM applied to the precluster model for $\alpha$ decay, we have shown that electromagnetic fields of strengths (\ref{Eeff}) exceeding the critical field of quantum electrodynamics are needed to considerably alter the rate of nuclear $\alpha$ decay. Proposals have been put forward to overcome the critical field limit by collisions of ultra relativistic nuclear beams with laser pulses of sub-critical intensity \cite{dipiazza-rmp12}. Furthermore, simulations indicate that the QED cascades which are expected at ultra-high electromagnetic fields \cite{kirk-prl08,fedotov-prl10} and put a conceptual limit to extreme laser intensities due to the laser energy depletion, may be controlable \cite{tamburini-sr17}. 
However, these methods remain far beyond the present experimental capabilities and many orders of magnitude in laser intensity are yet to be conquered. 
We conclude that laser-assisted $\alpha$ decay is practically out of reach and refute recent optimistic claims on the efficiency of this process and prospects for nuclear waste recycling.


We would like to acknowledge discussions with M. Ivanov, C. H. Keitel and G. R\"opke. 
SVP acknowledges financial support from the Ministry of Education and Science of the Russian Federation (grant No. 3.1659.2017/4.6).


\begin{thebibliography}{999}

\bibitem{ELI} J-P. Chambaret, O. Chekhlov, G. Cheriaux \etal, Extreme light infrastructure: laser architecture and major challenges, in ``Solid State Lasers and Amplifiers IV, and High-Power Lasers'' {\bf 7721}, 77211D (2010).

\bibitem{J-Karen} A. S. Pirozhkov, Y. Fukuda, M. Nishiuchi, H. Kiriyama, A. Sagisaka, K. Ogura, M. Mori, M. Kishimoto, H. Sakaki, N. P. Dover, K. Kondo, N. Nakanii, K. Huang, M. Kanasaki, K. Kondo and M. Kando, {\em Opt. Exp} {\bf 25}, 20486 (2017).

\bibitem{SULF} Z. Guo, L. Yu, J. Wang, C. Wang, Y. Liu, Z. Gan, W. Li, Y. Leng, X. Liang and R. Li, {\em Opt. Exp} {\bf 26}, 26776 (2017).

\bibitem{P3-ELI} S. Weber, et. al, {\em Matter Radiat. Extremes} {\bf 2}, 149 (2017). 

\bibitem{light-f} E. Cartlidge, {\em Science} {\bf 359}, 382 (2018).

\bibitem{XCELS} A.V. Bashinov, A.A. Gonoskov, A.V. Kim, G. Mourou and A.M. Sergeev, {\em  Eur. Phys. J.  Spec. Top.} {\bf 223}, 1105 (2014).

\bibitem{bulanov-rmp09} G. Mourou, T. Tajima and S.V. Bulanov,  \rmp {\bf 78}, 309 (2009).

\bibitem{dipiazza-rmp12} A. Di Piazza, C. M\"{u}ller, C.Z. Hatsagortsyan and C.H. Keitel, \rmp {\bf 84}, 1177 (2012).

\bibitem{fedotov-cp15} N.B. Narozhny and A.M. Fedotov,  Contemp. Phys. {\bf 56}, 249 (2015).


\bibitem{hector} H.~M. Casta\~neda Cortes, C. M\"uller, C.~H. Keitel and  A. P\'alffy, Phys. Lett B {\bf 723}, 401 (2013).

\bibitem{misicu-jpg13} S. Misicu and M. Rizea, J. Phys. G: Nucl. Part. Phys. {\bf 40}, 095101 (2013).

\bibitem{misicu-degruyter16} S. Misicu and M. Rizea, Open. Phys. {\bf 14}, 81 (De Gruyter, 2016); DOI 10.1515/phys-2016-0001. 

\bibitem{delion-prl17} D.~S. Delion and S.~A. Ghinescu, \prl {\bf 119}, 202501 (2017).

\bibitem{kis-jpg18} D.P. Kis and R. Szilvasi, J. Phys. G: Nucl. Part. Phys. {\bf 45}, 045103 (2018).

\bibitem{bai} D. Bai, D. Deng, Z. Ren, Nucl. Phys. A {\bf 976}, 23 (2018). 

\bibitem{qi-prc19} J. Qi, T. Li, R. Xu, L. Fu, and X. Wang, Phys. Rev. C {\bf 99} (2019).

\bibitem{BlattWeisskopf}  J. M. Blatt and V. F. Weisskopf, {\em Theoretical Nuclear Physics}, (Dover Publications, Inc.) 1991.


\bibitem{radioactive_decays} M. Pf\"utzner, M. Karny, L. V. Grigorenko and  K. Riisager, Rev.~Mod.~Phys. 84, 567 (2012).

\bibitem{Gamow1928} G. Gamow, Z. Phys. A 51, 204 (1928).

\bibitem{ConGur1928} R. W. Gurney and E. U. Condon, Phys. Rep. 33, 127 (1929).

\bibitem{GurKalb1987}  S. A. Gurvitz and G. Kalbermann, Phys. Rev. Lett. {\bf  59} 262 (1987).

\bibitem{XFEL} T. Ishikawa, Phil. Trans. Royal Soc. A (2019): doi/10.1098/rsta.2018.0231.

\bibitem{sauter} F.Z. Sauter, Phys. {\bf 69} 742 (1931); Z. Phys. {\bf 73} 547 (1932).

\bibitem{heisenberg} W. Heisenberg, H.Z. Euler, Phys. {\bf 98} 714 (1936).

\bibitem{schwinger} J. Schwinger, Phys. Rev. {\bf 82} 664 (1951)

\bibitem{PW-lasers} C. Danson, D. Hillier, N. Hopps and D. Neely, {\em High Power Laser Sci. Eng.} {\bf 3}, e3 (2015).

\bibitem{yan-oe08}  V. Yanovsky, V. Chvykov, G. Kalinchenko, P. Rousseau, T. Planchon, T. Matsuoka, A. Maksimchuk, J. Nees, G. Cheriaux, G. Mourou and K. Krushelnick, {\em Opt. Exp.} {\bf 16}, 2109 (2008).

\bibitem{ritus} V.I. Ritus, Zh. Eksp. Teor. Fiz. {\bf 56}, 986 (1969) [Sov. Phys. JETP {\bf 29}, 532 (1969)].

\bibitem{becker} W. Becker \etal, Phys. Lett. A {\bf 94}, 131 (1983).

\bibitem{rev} Chong Qi, Roberto Liotta and Ramon Wyss, Prog. Part. Nucl. Phys. {\bf 105}, 214 (2019).

\bibitem{LandauBook} L.D. Landau, E.M. Lifshitz, {\em Quantum mechanics. Nonrelativistic theory}, 2nd Ed., Pergamon Press, 1965.

\bibitem{popov-jetpl91} V.D. Mus, V.S. Popov, A.V. Sergeev, JETP Lett. {\bf 53}, 455 (1991).

\bibitem{Fermi} E. Fermi, {\em Nuclear Physics}, Chicago, 1950.


\bibitem{Buck} B. Buck, A. C. Merchant and S. M. Perez, Phys. Rev. Lett. 65, 2975 (1990), J. Phys. G: Nucl. Part. Phys. 17, 1223 (1991).

\bibitem{Royer} H. F. Zhang and G. Royer, Phys. Rev. C 77, 054318 (2008).

\bibitem{ensdf} Evaluated Nuclear Structure Data Files, https://www.nndc.bnl.gov/ensdf/ (2019).

\bibitem{popov-usp04} V.S. Popov, Phys. Usp. {\bf 49}, 855 (2004).

\bibitem{popov-itm} V. S. Popov, Phys. At. Nuclei {\bf 68}, 686 (2005).

\bibitem{lat2011} H.~M. Casta\~neda Cortes, S. V. Popruzhenko, D. Bauer and  A. P\'alffy, New J. Phys.{\bf 13}, 063007 (2011).

\bibitem{poprz-pra19} M.F. Ciappina, S.V. Popruzhenko, S.V. Bulanov \etal, Phys. Rev. A {\bf 99}, 043405 (2019).

\bibitem{Eberly} J. Javanainen, J. H. Eberly and Q. C. Su, Phys. Rev. A 38, 3430 (1988), Q. Su and J. H. Eberly, Phys. Rev. A 44, 5997 (1991). 

\bibitem{misha} We are thankful to M. Ivanov for this illuminating argument which points out once more the physical inconsistency of Refs.~\cite{delion-prl17,kis-jpg18,bai}.

\bibitem{henneberger} W.~C. Henneberger, \prl {\bf 21}, 838 (1968).

\bibitem{fedorov} M.~V. Fedorov, {\it Atomic and Free Electrons In A Strong Light Field}, (World
Scientific, 1997), p. 362; A.M. Popov, O.V. Tikhonova, E.A. Volkova, J. Phys. B: At. Mol. Opt. Phys. {\bf 32}, 3331 (1999).

\bibitem{kirk-prl08} A. R. Bell and J. G. Kirk, \prl {\bf 101} 200403 (2008).

\bibitem{fedotov-prl10} A.M. Fedotov, N.B. Narozhny, G. Korn and G. Mourou, \prl {\bf 105}, 080402 (2010).

\bibitem{tamburini-sr17} M. Tamburini, A. DiPiazza and Ch. Keitel, Sci. Rep. {\bf 7}, 5694 (2017).

\end{thebibliography}
\end{document}